\documentclass[aps,prl,twocolumn,showpacs,preprintnumbers,amsmath,amssymb,superscriptaddress,floatfix,nofootinbib]{revtex4-2}
\usepackage{amssymb}
\usepackage{color}
\usepackage{graphicx}
\usepackage{threeparttable}
\usepackage{upgreek}
\usepackage{lineno}
\usepackage[colorlinks=true, linkcolor=blue, citecolor=blue, urlcolor=blue]{hyperref}
\usepackage[none]{hyphenat}
\usepackage{float}
\usepackage{setspace}
\usepackage[T1]{fontenc}
\usepackage{titlesec}
\begin{document}
\title{Simultaneous $\alpha\beta$ Decay: A New Mode of Nuclear Instability}
\author{W.\ Q.\ Zhang}
\email{zwqzwq@impcas.ac.cn}
\affiliation{State Key Laboratory of Heavy Ion Science and Technology, Institute of Modern Physics, Chinese Academy of Sciences, Lanzhou 730000, China}
\affiliation{University of Chinese Academy of Sciences, Beijing 100049, China}
\author{C.\ Qi}
\email{chongq@kth.se}
\affiliation{KTH, Alba Nova University Center, SE-10691 Stockholm, Sweden}
\date{\today}
	
\begin{abstract}
We propose simultaneous $\alpha\beta$ decay as a novel mode of nuclear instability that involves the strong and weak interactions in a single quantum transition. We develop a theoretical framework to predict its branching ratios and $\alpha$-energy spectra, establishing exclusive and inclusive criteria based on whether the individual $\alpha$ and $\beta$ channels are closed or open. A global survey of the nuclear chart identifies five exclusive $\alpha\beta^-$ candidates, all predicted to be experimentally inaccessible, and ranks the leading inclusive candidates for both the $\alpha\beta^-$ and $\alpha\beta^+$ modes. Remarkably, six of the top $\alpha\beta^+$ candidates coincide with known $\beta$-delayed-$\alpha$ precursors. The observed $\alpha$ spectra are naturally accounted for by simultaneous $\alpha\beta^+$ emission  without additional assumptions, indicating a substantial and possibly dominant direct component. Our findings establish simultaneous $\alpha\beta$ decay as a distinct radioactive process and a sensitive probe of the interplay between the strong and weak interactions.
\end{abstract}	
\maketitle
	
Radioactive decay is a fundamental quantum process driving unstable nuclei toward energetically favorable configurations via the strong, electromagnetic, or weak interaction. The conventional $\alpha$, $\beta$, and $\gamma$ decay modes have long formed the cornerstones of our understanding of nuclear instability. In recent decades, exotic multiparticle decay modes involving the simultaneous emission of two or more identical particles, including the $2\nu 2\beta$ and $0\nu 2\beta$ decays driven by the weak interaction~\cite{RevModPhys.80.481,Gomez-Cadenas2023}, electromagnetic $2\gamma$ decay~\cite{PhysRevC.8.216,PhysRevLett.53.1897,PhysRevLett.133.022502}, and hadronic $2n$~\cite{PhysRevLett.110.152501}, $2p$~\cite{Blank2008,PFUTZNER2023104050,PhysRevLett.94.232501} and $2\alpha$~\cite{PhysRevLett.127.012501,DENISOV2022137569,jxqh-6gpj} decays, have been systematically investigated, greatly enriching the landscape of known radioactive processes. These rare decay modes can serve as sensitive probes of few-body correlations~\cite{RevModPhys.89.035006}, shell effects~\cite{PhysRevLett.107.062503}, and the limits of nuclear stability~\cite{Ravlic2023}. Yet, despite substantial attention to multiparticle extensions of single established decay modes, simultaneous transitions involving two distinct decay modes have rarely been considered, with only a few precedents in radiative $\beta$ decay~\cite{PhysRev.93.518,Nico2006} and virtual $\beta\gamma$ transitions~\cite{PhysRev.128.1776,PhysRevC.92.044305}. This raises a fundamental question: if two decay modes of a given nucleus are energetically forbidden or disfavored individually but allowed simultaneously, can the nucleus decay through the combined process as a single quantum transition?

\begin{figure*}[htb]
\begin{center}
\includegraphics[width=0.96\textwidth]{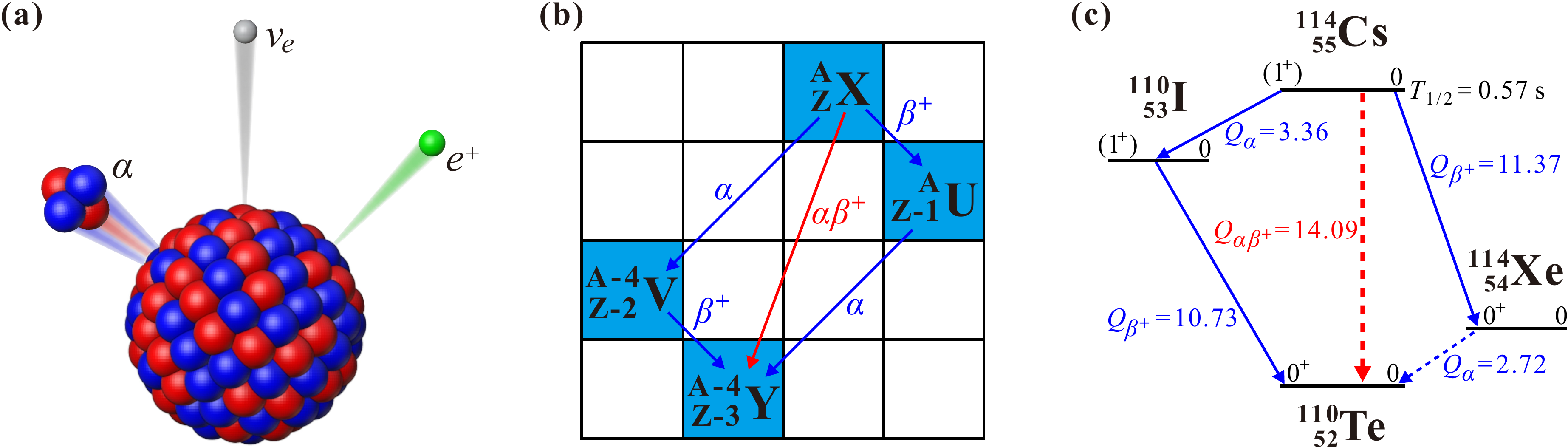}
\caption{(a) Schematic illustration of simultaneous $\alpha\beta^+$ decay. (b) The $\alpha$, $\beta^+$, and $\alpha\beta^+$ decay paths on the nuclear chart. (c) Decay modes of the leading $\alpha\beta^+$ candidate $^{114}$Cs, where solid and dashed arrows indicate observed and unobserved transitions, respectively, and the $Q$ values are given in MeV.}
\label{fig1}
\end{center}
\end{figure*}

In this work, we investigate the possibility of simultaneous $\alpha\beta$ decay as a distinct mode of nuclear instability. It can be classified into three types: $\alpha\beta^-$, $\alpha\beta^+$, and $\alpha$EC. Taking the $\alpha\beta^+$ type as an example, Figs.~\ref{fig1}(a) and \ref{fig1}(b) illustrate the decay process and the corresponding path on the nuclear chart, respectively. Unlike hypothetical baryon-number-violating proton decay~\cite{NATH2007191} and lepton-number-violating $0\nu 2\beta$ decay~\cite{RevModPhys.80.481}, simultaneous $\alpha\beta$ decay does not violate any known exact conservation law and is therefore allowed in principle. It provides a rare example in which weak-interaction operators act directly on cluster configurations, offering a unique testing ground for the interplay between the strong and weak interactions. Through detailed balance, the same matrix element formally defines a weak-force-assisted $\alpha$ capture process, although its astrophysical impact remains to be explored. 

We focus our theoretical framework on the $\alpha\beta^-$ and $\alpha\beta^+$ modes, where the emitted leptons provide a well-defined leptonic phase space. The simultaneous $\alpha\beta$ transition connects an initial nucleus to a final state containing an $\alpha$ particle, a daughter nucleus, and leptons. Neither the $\alpha$-emission operator $\hat{O}_\alpha$ nor the weak operator $\hat{O}_\beta$ alone can connect these states, since each changes only part of the final-state configuration. Consequently, the first-order transition amplitude vanishes. The leading non-zero contribution arises from the combined action of the two interactions and appears at second order in perturbation theory. By inserting complete sets of virtual intermediate states $\{|n\rangle\}$ and $\{|m\rangle\}$, the transition amplitude reads:
\begin{equation}
\label{eq1}
\begin{aligned}
\mathcal{M}_{fi} ={}&
\sum_n
\frac{
\langle {}^{A-4}_{Z_d}\mathrm{Y};f
\|\hat{O}_\beta\|n\rangle
\langle n\|\hat{O}_\alpha\|
{}^A_Z\mathrm{X};i\rangle
}{
E_i-E_n-m_\alpha c^2-E_\alpha
}\\
&+
\sum_m
\frac{
\langle {}^{A-4}_{Z_d}\mathrm{Y};f
\|\hat{O}_\alpha\|m\rangle
\langle m\|\hat{O}_\beta\|
{}^A_Z\mathrm{X};i\rangle
}{
E_i-E_m-m_ec^2-E_{\rm lept}
},
\end{aligned}
\end{equation}
where $i$ and $f$ denote the initial and final nuclear states, respectively, and $E_i$, $E_n$, and $E_m$ denote the energies of the corresponding nuclear states. Here $E_\alpha$ is the sum of the kinetic energies of the emitted $\alpha$ particle and final daughter nucleus, and $E_{\rm lept}$ is the total kinetic energy of the emitted leptons. Both are defined in the parent-nucleus rest frame, so that $E_\alpha+E_{\rm lept}=Q_{\alpha\beta}$. The weak transition operator $\hat{O}_{\beta}$ is dimensionless, while the $\alpha$-cluster emission operator $\hat{O}_{\alpha}$ is related to the microscopic formation amplitude $F(R)$ by the R-matrix relation $\langle \hat{O}_{\alpha}\rangle^2 = \frac{\hbar^2 R}{\mu} F(R)^2$~\cite{Qi2019,QI2021136373}. As a result, $\mathcal{M}_{fi}$ assumes the dimension ${\rm MeV}^{-1/2}$. No exact orthogonality relation makes every term in the virtual-state sum vanish, nor does any symmetry enforce an exact cancellation among all virtual-state contributions; hence, \(\mathcal{M}_{fi}\) is generically nonzero.

The simultaneous $\alpha\beta$-decay width can be derived from Fermi's Golden rule by integrating over the four-body final-state phase space, which reduces to an integration over $E_\alpha$:
\begin{equation}
\label{eq2}
    \Gamma_{\alpha\beta}
    =
    \frac{\hbar\ln 2}{T_{1/2}^{\alpha\beta}}
    =
    \int_0^{Q_{\alpha\beta}}
    \frac{d\Gamma_{\alpha\beta}}{dE_\alpha}\,dE_\alpha ,
\end{equation}
where
\begin{equation}
\label{eq3}
    \frac{d\Gamma_{\alpha\beta}}{dE_\alpha}
    =
    C_0
    \left|\mathcal{M}_{fi}\right|^2
    P_\alpha(E_\alpha,l_\alpha)
    f\left(Z_d,W_0\right).
\end{equation}
Here, $P_\alpha(E_\alpha,l_\alpha)$ is the $\alpha$-barrier penetrability at energy $E_\alpha$ and angular momentum $l_\alpha$. The value of $l_\alpha$ is constrained by the angular-momentum coupling between the emitted $\alpha$ particle and the leptonic system, and we use the lowest compatible value in the calculations. $f(Z_d,W_0)$ is the dimensionless Fermi integral~\cite{Behrens1982} for a $\beta$ transition into the residual daughter charge $Z_d$, with the endpoint energy $W_0=1+E_{\rm lept}/(m_ec^2)$. The prefactor $C_0$  accounts for the remaining kinematic and weak-coupling normalization factors. It is derived to be $C_0=\frac{G_V^2(m_ec^2)^5}{4\pi^4(\hbar c)^6}$, where $G_V$ is the vector weak coupling constant (dimension ${\rm MeV}\,{\rm fm}^3$). $C_0$ has the dimension of energy (${\rm MeV}$). The differential decay width $d\Gamma_{\alpha\beta}/dE_\alpha$ is therefore dimensionless. The penetrability $P_\alpha(E_\alpha,l_\alpha)$ is calculated with the $l$-dependent Coulomb-Hankel function $H^+_{l}$, following the standard charged-particle-emission formalism:
\begin{equation}
\label{eq4}
    P_\alpha(E_\alpha,l_\alpha)
    =
    \rho\left|H^+_{l}(\eta,\rho)\right|^{-2},
\end{equation}
with the notation defined as in Ref.~\cite{Qi2019}. 

The dramatically different energy dependences of $P_\alpha(E_\alpha,l_\alpha)$ and $f(Z_d,W_0)$ make the integrand $g(E_\alpha,l_\alpha)\equiv P_\alpha(E_\alpha,l_\alpha)f(Z_d,W_0)$ in Eq.~\eqref{eq2} strongly peaked at an intermediate energy $E_\alpha^*$. With $|\mathcal{M}_{fi}|^2$ treated as constant near this energy, the peak position is determined from the saddle-point condition:
\begin{equation}
\begin{aligned}
\label{eq5}
    &\left.
    \frac{d}{dE_\alpha}
    \ln P_\alpha(E_\alpha,l_\alpha)
    \right|_{E_\alpha=E_\alpha^*}  \\
    &\qquad =
    \left.
    \frac{1}{m_ec^2}
    \frac{d}{dW_0}
    \ln f(Z_d,W_0)
    \right|_{W_0=W_0^*},
\end{aligned}
\end{equation}
where $W_0^*=1+E_{\rm lept}^*/(m_ec^2)$ and $E_{\rm lept}^*=Q_{\alpha\beta}-E_\alpha^*$. The peak width $\sigma$ is obtained from the curvature of the integrand at the saddle point:
\begin{equation}
\label{eq6}
    \sigma
    =
    \left[
    -
    \left.
    \frac{d^2}{dE_\alpha^2}
    \ln g(E_\alpha,l_\alpha)
    \right|_{E_\alpha=E_\alpha^*}
    \right]^{-1/2}.
\end{equation}
Then, by expanding $g(E_\alpha,l_\alpha)$ around $E_\alpha^*$, the decay width $\Gamma_{\alpha\beta}$ can be approximated by
\begin{equation}
\label{eq7}
    \Gamma_{\alpha\beta}
    \simeq
    C_0\left|\mathcal{M}_{fi}\right|^2
    \sqrt{2\pi}\,\sigma
    P_\alpha(E_\alpha^*,l_\alpha)\,f(Z_d,W_0^*) .
\end{equation}
We define the effective reduced width $\gamma_{\alpha\beta}$ of simultaneous $\alpha\beta$ decay, normalized by the explicit $\alpha$-barrier penetrability and leptonic phase-space factor, as
\begin{equation}
\label{eq8}
    \gamma_{\alpha\beta}
    \equiv
    C_0\left|\mathcal{M}_{fi}\right|^2
    \sqrt{2\pi}\,\sigma
   \simeq
    \frac{
    \Gamma_{\alpha\beta}
    }{
    P_\alpha(E_\alpha^*,l_\alpha)\,f(Z_d,W_0^*)
    } .
\end{equation}

We now apply this framework to a global search for simultaneous $\alpha\beta$-decay candidates. It is essential to distinguish the exclusive and inclusive cases of simultaneous $\alpha\beta$ decay. The former requires the combined channel to be open and the individual $\alpha$ and $\beta$ channels to be closed, whereas the latter requires only the simultaneous channel to be open. We first consider the exclusive $\alpha\beta^\pm$ case, defined by $Q_{\alpha\beta} > 0$, $Q_{\alpha} < 0$, and $Q_{\beta} < 0$ (where $Q_{\beta}$ denotes the $Q$ values for all $\beta^\pm$ and EC decays), which avoids competition with ordinary decay modes. The $Q_{\alpha\beta}$ values are calculated from available atomic mass data~\cite{CPC}. This search yields only five nuclei satisfying the exclusive criterion for simultaneous $\alpha\beta$ decay: $^{150}$Nd, $^{157}$Gd, $^{159}$Tb, $^{163}$Dy, and $^{204}$Hg. All five candidates belong to the $\alpha\beta^-$ type and are naturally occurring isotopes. For the most favorable exclusive candidate, $^{163}$Dy, the released decay energy is only $Q_{\alpha\beta^-}=727$ keV. Under the most optimistic assumption that the full decay energy is carried by the emitted $\alpha$ particle, the universal decay law~\cite{Qi_PRL2009,Qi_PRC2009} yields a lower limit for the half-life on the order of $10^{67}$ yr. Such an ultralong timescale lies far beyond the reach of existing rare-event detection technologies~\cite{PhysRevD.102.112011,v4c6-h6l6,PhysRevLett.130.062501,ZHANG2026}.

We therefore move beyond the exclusive energetic condition and search for inclusive candidates, which require only $Q_{\alpha\beta}>0$ and may have appreciable simultaneous $\alpha\beta$-decay branching ratios.
For the purpose of identifying candidates with relatively large branching ratios, we assume that $\gamma_{\alpha\beta}$ is the same for all candidates of a given mode. Under this assumption, Eq.~\eqref{eq8} gives the relative $\alpha\beta$ branching ratio between two candidate nuclei as
\begin{equation}
\label{eq9}
\frac{b_{\alpha\beta}(i)}{b_{\alpha\beta}(j)}
=
\frac{T_{1/2}^{\rm tot}(i)}{T_{1/2}^{\rm tot}(j)}
\frac{
P_\alpha(E_{\alpha,i}^*,l_{\alpha,i})\,f(Z_{d,i},W_{0,i}^*)
}{
P_\alpha(E_{\alpha,j}^*,l_{\alpha,j})\,f(Z_{d,j},W_{0,j}^*)
},
\end{equation}
where $i$ and $j$ label the two nuclei to be compared, and $T_{1/2}^{\rm tot}$ denotes their experimental half-lives. Based on Eq.~\eqref{eq9}, we rank the inclusive candidates separately for the simultaneous $\alpha\beta^-$ and $\alpha\beta^+$ decays, restricting the search to nuclei with $A\geq100$, where the present barrier-penetration treatment of $\alpha$ emission is most applicable, and considering only ground-state-to-ground-state transitions. The top ten candidates for the two modes are summarized in Tables~\ref{tab1} and \ref{tab2}, respectively. 

For simultaneous $\alpha\beta^-$ decay, as listed in Table~\ref{tab1}, all the top ten inclusive candidates are neutron-rich nuclei and lie close to the northeast side of the doubly magic nuclei $^{132}$Sn and $^{208}$Pb. Notably, the leading two candidates $^{136}$Sb and $^{212}$Bi decay into the two doubly magic nuclei, respectively, with the predicted $\alpha\beta^-$ branching ratio of $^{136}$Sb approximately two orders of magnitude larger than that of $^{212}$Bi.

\begin{table}[!t]
\caption{Top ten candidate nuclei for simultaneous $\alpha\beta^-$ decay, ranked by the calculated branching ratios normalized to the largest value, 
$b_{136}$=$b_{\alpha\beta^-}(^{136}{\rm Sb})$. The peak energy $E_\alpha^{{\rm k}*}$ and width $\sigma^{\rm lab}$ are given for the kinetic energy of the $\alpha$ particle.} 
\label{tab1}
\begin{ruledtabular}
\begin{tabular}{cccccc}
Parent & Daughter & 
$Q_{\alpha\beta^-}$ &
$E_\alpha^{{\rm k}*}$ &
$\sigma^{\rm lab}$ &
$b_{\alpha\beta^-}/b_{136}$ \\
nucleus & nucleus &
(MeV) & (MeV) & (MeV) & \\
\hline
$^{136}$Sb & $^{132}$Sn & $9.10$  & $7.80$ & $0.50$ & $1$ \\
$^{212}$Bi & $^{208}$Pb & $10.70$ & $9.71$ & $0.41$ & $9.83\times10^{-3}$ \\
$^{137}$Sb & $^{133}$Sn & $7.88$  & $6.86$ & $0.39$ & $3.68\times10^{-3}$ \\
$^{138}$I  & $^{134}$Te & $7.62$  & $6.69$ & $0.36$ & $3.04\times10^{-3}$ \\
$^{214}$Bi & $^{210}$Pb & $10.59$ & $9.62$ & $0.40$ & $2.15\times10^{-3}$ \\
$^{140}$I  & $^{136}$Te & $7.88$  & $6.91$ & $0.38$ & $4.91\times10^{-4}$ \\
$^{137}$Te & $^{133}$Sb & $6.68$  & $5.91$ & $0.30$ & $1.75\times10^{-5}$ \\
$^{216}$Bi & $^{212}$Pb & $10.49$ & $9.55$ & $0.39$ & $1.67\times10^{-5}$ \\
$^{213}$Bi & $^{209}$Pb & $9.45$  & $8.64$ & $0.33$ & $1.46\times10^{-5}$ \\
$^{140}$Cs & $^{136}$Xe & $6.44$  & $5.75$ & $0.26$ & $6.46\times10^{-6}$ \\
\hline
\end{tabular}
\end{ruledtabular}
\end{table}

\begin{table}[!t]
\caption{Top ten candidate nuclei for simultaneous $\alpha\beta^+$ decay, given as in Table~\ref{tab1}, with branching ratios normalized to $b_{114}$=$b_{\alpha\beta^+}(^{114}{\rm Cs})$.}
\label{tab2}
\begin{ruledtabular}
\begin{tabular}{cccccc}
Parent & Daughter & 
$Q_{\alpha\beta^+}$ &
$E_\alpha^{{\rm k}*}$ &
$\sigma^{\rm lab}$ &
$b_{\alpha\beta^+}/b_{114}$ \\
nucleus & nucleus &
(MeV) & (MeV) & (MeV) & \\
\hline
$^{114}$Cs$^\dagger$ & $^{110}$Te & $14.09$ & $11.20$ & $0.91$ & $1$ \\
$^{110}$I$^\dagger$  & $^{106}$Sn & $13.44$ & $10.67$ & $0.87$ & $8.09\times10^{-1}$ \\
$^{112}$I$^\dagger$  & $^{108}$Sn & $11.56$ & $9.41$  & $0.69$ & $7.07\times10^{-2}$ \\
$^{109}$Xe           & $^{105}$Sb & $14.40$ & $11.33$ & $0.95$ & $5.32\times10^{-2}$ \\
$^{115}$Ba           & $^{111}$I  & $12.59$ & $10.24$ & $0.75$ & $2.11\times10^{-2}$ \\
$^{116}$Cs$^\dagger$ & $^{112}$Te & $12.08$ & $9.85$  & $0.72$ & $1.88\times10^{-2}$ \\
$^{113}$Xe$^\dagger$ & $^{109}$Sb & $10.60$ & $8.76$  & $0.60$ & $2.28\times10^{-3}$ \\
$^{118}$Cs$^\dagger$ & $^{114}$Te & $10.03$ & $8.38$  & $0.54$ & $1.01\times10^{-3}$ \\
$^{170}$Ir           & $^{166}$W  & $15.26$ & $12.97$ & $0.78$ & $9.78\times10^{-4}$ \\
$^{111}$I            & $^{107}$Sn & $10.11$ & $8.36$  & $0.56$ & $8.53\times10^{-4}$ \\
\hline
\multicolumn{6}{l}{\footnotesize $^\dagger$ Reported $\beta$-delayed-$\alpha$ precursors.}
\end{tabular}
\end{ruledtabular}
\end{table}

For simultaneous $\alpha\beta^+$ decay, all ten candidates are neutron-deficient nuclei, with nine lying near and northeast of the doubly magic nucleus $^{100}$Sn, as listed in Table~\ref{tab2}. Their $Q_{\alpha\beta^+}$ values are generally larger than the $Q_{\alpha\beta^-}$ values of the $\alpha\beta^-$ candidates (Table~\ref{tab1}), reflecting the fact that the neutron-deficient region favors both $\alpha$ and $\beta^+$ decays, whereas the neutron-rich region favors $\beta^-$ decay only. This renders the $\alpha\beta^+$ mode more promising for experimental searches. The decay modes of the most promising $\alpha\beta^+$ candidate $^{114}$Cs are depicted in Fig.~\ref{fig1}(c).

Remarkably, six of the ten $\alpha\beta^+$ candidates, namely $^{110,112}$I, $^{113}$Xe, and $^{114,116,118}$Cs, have already been reported experimentally as $\beta$-delayed-$\alpha$ precursors~\cite{NNDC,HORNSHOJ1975,BOGDANOV1977,HAGBERG1978,DAURIA1978,Roeckl1980,TIDE1982,TIDE1985}. For five of them, the corresponding $\alpha$ spectra are available, all exhibiting Gaussian-like peaks with MeV-scale widths~\cite{BOGDANOV1977,HAGBERG1978,DAURIA1978,Roeckl1980,TIDE1982,TIDE1985,Dey_PRL2026}. However, statistical-model calculations~\cite{HORNSHOJ1972609,Jonson1976} did not reproduce the measured $\alpha$ spectra and branching ratios; the predicted widths are larger and the predicted branching ratios substantially smaller than observed~\cite{TIDE1982,TIDE1985}. Tidemand-Petersson \textit{et al.}~\cite{TIDE1982} showed that agreement with the data required three phenomenological modifications in the statistical model: 
(1) an enhanced level density, (2) a Gaussian resonance in the $\beta$-strength function, and (3) $\gamma$-decay widths increased by a factor of ten. 

In addition, within the conventional $\beta$-delayed-$\alpha$ picture, if each observed Gaussian-like $\alpha$ spectrum is assigned to a single $\alpha$-emitting intermediate state, its MeV-scale width corresponds to a half-life of order $10^{-22}$~s. For these $\alpha$ decays with energies around $9$~MeV, the corresponding $\alpha$-decay reduced width $\delta^2$~\cite{Rasmussen1959} can be deduced to reach the GeV scale, far beyond physically reasonable values. Consequently, the $\beta$-delayed-$\alpha$ interpretation must invoke a smooth superposition of a dense set of unresolved $\beta$-fed $\alpha$-emitting states. However, a recent experimental study of $^{116}$Cs~\cite{Dey_PRL2026} assigns the $\alpha$ events to a single $\alpha$-decaying state with a width of 2431 keV in the $\beta$-decay daughter nucleus $^{116}$Xe. Moreover, at the same excitation energy, a much narrower proton-emitting structure with a width of $111$~keV is resolved in the delayed-proton spectrum, whereas no corresponding structure is observed in the delayed-$\alpha$ spectrum.

These difficulties motivate a comparison between the $\beta$-delayed-$\alpha$ emission interpretation and the simultaneous $\alpha\beta^+$ decay interpretation. The essential difference between the two pictures lies in the decay mechanism. Conventional $\beta$-delayed-$\alpha$ emission is a sequential two-step process that requires the population of real excited states in the intermediate nucleus. By contrast, simultaneous $\alpha\beta^+$ decay proceeds, through virtual intermediate states, as a single quantum transition directly to the same four-body final state. Consequently, the simultaneous $\alpha\beta^+$ component may become increasingly competitive at large $Q_{\alpha\beta^+}$, where the decay energy can be shared continuously among the $\alpha$ particle and the leptons, allowing the $\alpha$ particle to carry enough energy to overcome the Coulomb barrier while still leaving sufficient energy for the leptonic phase space.

\begin{figure}[t]
\begin{center}
\includegraphics[width=0.48\textwidth]{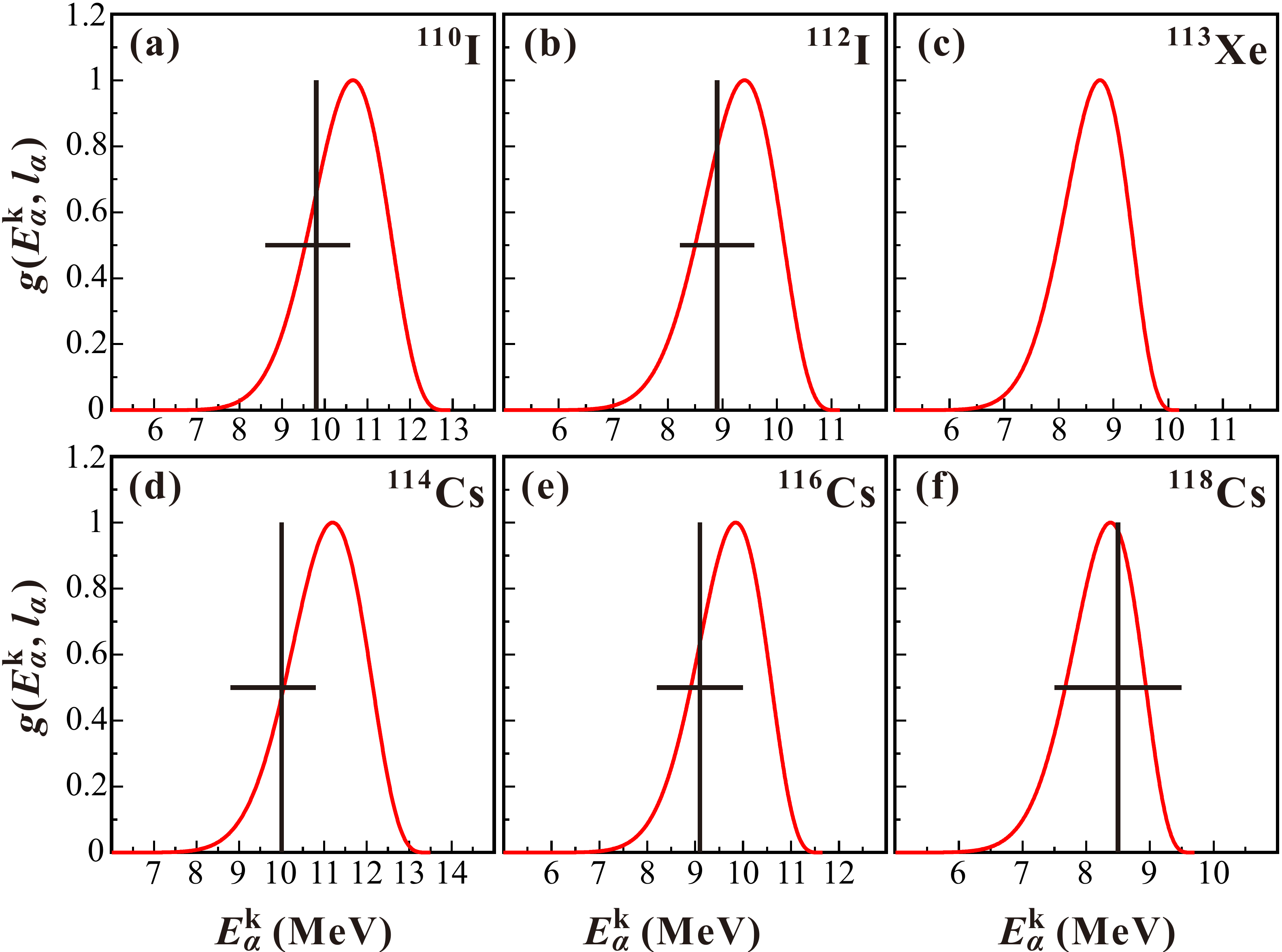}
\caption{The integrand $g(E_\alpha^{\rm k},l_\alpha)$ for simultaneous $\alpha\beta^+$ decay in $^{110,112}$I, $^{113}$Xe, and $^{114,116,118}$Cs, normalized to unity at the maximum. $E_\alpha^{\rm k}$ denotes the kinetic energy of the $\alpha$ particle. The black vertical and horizontal bars indicate, respectively, the peak energy and full width at half maximum (FWHM) of the experimental $\alpha$ spectra previously assigned to $\beta$-delayed-$\alpha$ emission, as estimated from the published figures in Refs.~\cite{BOGDANOV1977,HAGBERG1978,DAURIA1978,Roeckl1980,TIDE1982,TIDE1985,Dey_PRL2026}.}
\label{fig2}
\end{center}
\end{figure}

Nevertheless, in both cases the positron and $\alpha$ particle would appear experimentally ``simultaneous'', making it almost impossible to distinguish the two mechanisms through their time correlation. Similar ambiguities between direct and sequential mechanisms are known in few-body decays, where the intermediate paths within the interaction region are not themselves observables. There the mechanism must be inferred from final-state energy and momentum distributions and from resonance signatures~\cite{PhysRevLett.100.192501}. For the cases considered here, the distinction relies mainly on the comparison between the predicted and experimental $\alpha$-energy spectra. As mentioned above, reproducing the observed Gaussian-like $\alpha$-energy spectra within the $\beta$-delayed-$\alpha$ interpretation requires the introduction of a Gaussian resonance in the $\beta$-strength function~\cite{TIDE1982}. For simultaneous $\alpha\beta^+$ decay, the $\alpha$-energy distribution is determined solely by the integrand $g(E_\alpha,l_\alpha)$, in analogy with the Fermi integral that gives the shape of ordinary $\beta$ spectra. 

\begin{table}[!t]
\caption{Derived effective reduced widths $\gamma_{\alpha\beta^+}$ for the six reported $\beta$-delayed-$\alpha$ precursors under the hypothesis that the measured $\beta$-delayed-$\alpha$ branching ratios correspond to simultaneous $\alpha\beta^+$ branching ratios. The experimental data are taken from Refs.~\cite{NNDC,HORNSHOJ1975,BOGDANOV1977,HAGBERG1978,DAURIA1978,Roeckl1980,TIDE1982,TIDE1985,Dey_PRL2026}.}
\label{tab3}
\begin{ruledtabular}
\begin{tabular}{ccccc}
Parent &
$T_{1/2}^{\rm expt}$ &
$b_{\alpha\beta^+}$ &
$T_{1/2}^{\alpha\beta^+}$ &
$\gamma_{\alpha\beta^+}$ \\
nucleus &
(s) &
(\%) &
(s) &
(MeV) \\
\hline
$^{110}$I  & $0.66$ & $1.1$    & $6.0\times10^{1}$ & $6.2\times10^{-23}$ \\
$^{112}$I  & $3.34$ & $0.10$   & $3.3\times10^{3}$ & $6.5\times10^{-23}$ \\
$^{113}$Xe & $2.74$ & $0.007$  & $3.9\times10^{4}$ & $1.4\times10^{-22}$ \\
$^{114}$Cs & $0.57$ & $0.19$   & $3.0\times10^{2}$ & $8.7\times10^{-24}$ \\
$^{116}$Cs & $0.70$ & $0.049$  & $1.4\times10^{3}$ & $1.2\times10^{-22}$ \\
$^{118}$Cs & $14$   & $0.0024$ & $5.8\times10^{5}$ & $1.1\times10^{-22}$ \\
\hline
\end{tabular}
\end{ruledtabular}
\end{table}

Figure~\ref{fig2} displays the normalized $g(E_\alpha^{\rm k},l_\alpha)$ curves for the six $\alpha\beta^+$ candidates reported as $\beta$-delayed-$\alpha$ precursors and compares them with the available experimental $\alpha$ spectra~\cite{BOGDANOV1977,HAGBERG1978,DAURIA1978,Roeckl1980,TIDE1982,TIDE1985,Dey_PRL2026}. Notably, detector broadening~\cite{TIDE1985} is negligible compared with the observed spectral widths. The simultaneous-emission picture qualitatively accounts for the Gaussian-like peak shapes and widths of the observed $\alpha$ spectra without introducing any additional assumptions, although the predicted peak positions deviate from the experimental values by $\sim 1$~MeV for $^{110,112}$I and $^{114,116}$Cs. The Gaussian-like peak shapes arise naturally from the optimal energy sharing between the $\alpha$ particle and the leptons, with the differential decay width $d\Gamma_{\alpha\beta^+}/dE_\alpha$ maximized at the saddle-point energy.

Furthermore, if the reported $\beta$-delayed-$\alpha$ branching ratios for the six precursors~\cite{NNDC} are assumed to be simultaneous $\alpha\beta^+$ branching ratios, Eq.~\eqref{eq8} would give the corresponding effective reduced widths, as listed in Table~\ref{tab3}. The extracted $\gamma_{\alpha\beta^+}$ values are all of similar order of magnitude, lying in the range $10^{-24}$--$10^{-22}$ MeV, and such small values are consistent with the expectation for a higher-order process involving the weak interaction. These arguments indicate that there is no obvious contradiction between the simultaneous $\alpha\beta^+$ interpretation and the experimental data previously assigned to $\beta$-delayed-$\alpha$ emission.

This leads to a central question: were the $\beta$-$\alpha$ events in the above-mentioned nuclei genuine sequential $\beta$-delayed-$\alpha$ emission, or were they manifestations of simultaneous $\alpha\beta^+$ decay? Although no conclusive evidence can yet distinguish between the two interpretations, the simultaneous $\alpha\beta$-emission picture offers a more natural explanation of the Gaussian-like $\alpha$ spectra. Experimentally, in addition to reinvestigating the six known cases with modern decay spectroscopy, $^{170}$Ir is of particular importance (Table~\ref{tab2}). The simultaneous-emission calculation predicts an $\alpha$ peak near 13 MeV for $^{170}$Ir, substantially higher than in the known cases; a dedicated search for this high-energy structure would provide an independent test in a different nuclear region. A microscopic calculation of the transition amplitude $\mathcal{M}_{fi}$ in Eq.~\eqref{eq1} is essential for a rigorous assessment of simultaneous $\alpha\beta$ decay, and would allow absolute branching ratios to be predicted. Such a calculation would incorporate realistic $\alpha$-cluster formation amplitudes and weak-transition matrix elements, coherently sum over virtual intermediate states, and treat continuum coupling and the finite widths of near-resonant states. Given this complexity and the limited experimental information on excited states in these extremely neutron-deficient nuclei, such a calculation remains a considerable challenge.

In summary, we have proposed and formulated a theoretical description of simultaneous $\alpha\beta$ decay as a new mode of nuclear instability. Under the exclusive criterion, only five $\alpha\beta^-$ candidates were identified across the nuclear chart, all beyond experimental reach. We evaluated inclusive candidates with potentially accessible branching ratios, finding that the leading $\alpha\beta^-$ candidates cluster near the doubly magic nuclei $^{132}$Sn and $^{208}$Pb, while the leading $\alpha\beta^+$ candidates lie near the doubly magic nucleus $^{100}$Sn. Remarkably, six $\alpha\beta^+$ candidates are known $\beta$-delayed-$\alpha$ precursors. Their observed Gaussian-like $\alpha$ spectra are naturally reproduced by our simultaneous-emission framework without free parameters. This indicates that phenomena historically interpreted as sequential $\beta$-delayed-$\alpha$ decay may be dominated by, or at least include substantial contributions from, direct, simultaneous $\alpha\beta^+$ emission. Future high-precision decay spectroscopy and microscopic many-body calculations will be essential to verify this decay mode and further probe the limits of nuclear stability.

\vspace{1\baselineskip}
\textit{Acknowledgments}---W. Q. Zhang was supported by the National Natural Science Foundation of China (Grant No. 12405140) and the Natural Science Foundation of Gansu Province, China (Grant No. 24JRRA033).
\bibliographystyle{apsrev4-2}
\bibliography{refabdecay}
\end{document}